\documentclass[12pt]{article}
\pdfoutput=1

\usepackage{draft,hyperref,tikz,cancel,subfig,float}
\usepackage{multirow}
\usepackage{soul}
\usepackage[nosort]{cite}

\usetikzlibrary{snakes}
\usetikzlibrary{shapes.misc}

\newtheorem{innercustomcoro}{Corollary}
\newenvironment{customcoro}[1]
{\renewcommand\theinnercustomcoro{#1}\innercustomcoro}
  {\endinnercustomcoro}

\newtheorem{innercustomthm}{Theorem}
\newenvironment{customthm}[1]
  {\renewcommand\theinnercustomthm{#1}\innercustomthm}
  {\endinnercustomthm}
  
\newtheorem{innercustomlem}{Lemma}
\newenvironment{customlem}[1]
  {\renewcommand\theinnercustomlem{#1}\innercustomlem}
  {\endinnercustomlem}

\begin{document}

\begin{titlepage}

\begin{center}

\hfill \\
\hfill \\
\vskip 1cm

\title{5d and 6d SCFTs Have No Weak Coupling Limit
}

\author{Chi-Ming Chang}

\address{
Center for Quantum Mathematics and Physics (QMAP) \\
University of California, Davis, CA 95616, USA}

\email{wychang@ucdavis.edu}

\end{center}

\abstract{
We prove that there does not exist any weak coupling limit in the space of superconformal field theories in five and six dimensions, based on an analysis of the representation theory of the corresponding superconformal algebras. Holographically, this implies that superstring theories on ${\rm AdS}_6$ and ${\rm AdS}_7$ do not admit tensionless limits. Finally, we discuss the implications of our result on the existence of an action for coincident M5-branes.
} 

\vfill

\end{titlepage}

\eject


%

Low energy theories on coincident M5-branes play important roles in the study of supersymmetric field theories, holography and M-theory. They provide examples of interacting conformal field theories with maximal supersymmetry with the highest spacetime dimensions, {\it i.e.} ${\cal N}=(2,0)$ supersymmetry in six dimensions. Their compactification down to lower dimensions generates a plethora of quantum field theories and dualities \cite{Gaiotto:2009we,Gaiotto:2009hg}. The $(2,0)$ theories also provide some of the first examples of the AdS/CFT correspondence \cite{Maldacena:1997re}. Recently, they have become powerful tools that elucidate M-theory amplitudes beyond the eleven-dimensional supergravity limit \cite{Chester:2018aca,Chester:2018dga}.

Since the low energy action for a single M5-brane was constructed in \cite{Howe:1996yn,Pasti:1996vs,Perry:1996mk,Schwarz:1997mc,Pasti:1997gx,Bandos:1997ui,Aganagic:1997zq,Howe:1997fb,Bandos:1997gm,Cederwall:1997gg}, it has been a longstanding goal to construct a Lorentz-invariant supersymmetric action for multiple coincident M5-branes. Despite tremendous endeavors \cite{Bergshoeff:1996qm,Ho:2008nn,Ho:2008ve,Chu:2008qv,Chu:2009iv,Chen:2010jgb,Ho:2011ni,Bunster:2011qp,Samtleben:2011fj,Chu:2012um,Bonetti:2012st,Ko:2013dka,Ko:2015zsy,Ganor:2017rrz,Saemann:2017zpd,Ananth:2018pzx}, such an action has not been successfully constructed. There are various no-go theorems excluding certain forms of the action \cite{Nepomechie:1982rb,Henneaux:1995ts,Henneaux:1997ha,Bekaert:1999dp,Henneaux:1999ma,Bekaert:2000qx,Bekaert:2001wa,Chen:2010ny,Huang:2010rn,Czech:2011dk}, but whether such an action exists still remains inconclusive. However, in three dimensions, a similar attempt to construct an action for coincident M2-branes has proven successful. In the low energy limit, the theories are three-dimensional conformal field theories with ${\cal N}=8$ supersymmetry, whose action was constructed in \cite{Bagger:2006sk,Bagger:2007jr,Bagger:2007vi,Gustavsson:2008dy,Distler:2008mk,Lambert:2008et,Aharony:2008ug}. The action has a family of extensions, parameterized by a positive integer $k$ and describing $N$ M2-branes probing a ${\mathbb C}^4/{\mathbb Z}_k$ singularity. In the large $k$ limit, the theory becomes weakly coupled.   Assuming $N > 2$, for $k>2$, the theory has ${\cal N}=6$ supersymmetry; however, for $k=1,2$, the supersymmetry is enhanced by quantum effects to ${\cal N}=8$.

The supersymmetry enhancement in three dimensions has motivated the search of an action for $(2,0)$ superconformal field theories in six dimensions, but with only manifest $(1,0)$ supersymmetry (for example, see \cite{Samtleben:2011fj,Bergshoeff:1996qm}). One may wonder if there exists a sequence of $(1,0)$ superconformal field theories labeled by a positive integer $k$, such that for special values of $k$ the theories have enhanced $(2,0)$ supersymmetry, and in the large $k$ limit the theories become weakly coupled, with the sequence converging to a free theory at infinite $k$.\footnote{The low energy limit of M5-branes probing a ${\mathbb C}^2/{\mathbb Z}_k$ singularity was studied in \cite{Brunner:1997gf,Hanany:1997gh,Ferrara:1998vf,Gaiotto:2014lca,DelZotto:2014hpa}, but this setup does not lead to a free theory in the $k\to \infty$ limit.} If such a sequence exists, one can first write down an action for the theories in the weak coupling limit at large $k$.\footnote{Such a perturbative action can be constructed by adding interaction terms order by order in $k^{-1}$ to the action of the free theory.} Then suitable extrapolation of the weakly coupled action to small $k$ may give an action for the $(2,0)$ theories. We will point out that such a sequence is inconsistent with the representation theory of the corresponding superconformal algebra. This is the key observation of this paper.
 
\subsubsection*{Main theorem}
 
Superconformal algebras in various dimensions are famously classified by Nahm \cite{Nahm:1977tg}. In six dimensions, there is an infinite family of superconformal algebras labeled by a positive integer $\cal N$, often denoted $({\cal N},0)$.\footnote{In these algebras, all Poincar\'{e} $Q$-supercharges have the same chirality.} The unitary superconformal multiplets (unitary irreducible representations of the superconformal algebra) in six dimension are classified in \cite{Buican:2016hpb,Cordova:2016emh}. Many strong constraints on the space of $({\cal N},0)$ theories can be derived from the representation theory of the superconformal algebra alone. For example, there is no local quantum field theory with $({\cal N}>2,0)$ supersymmetry because there is no superconformal multiplet that could contain a stress tensor \cite{Cordova:2016emh}. Unitarity further restricts the space of superconformal field theories to be discrete, since a continuous direction would imply the existence of a supersymmetry-preserving marginal operator, which cannot reside in a unitary superconformal multiplet \cite{Louis:2015mka,Cordova:2016xhm}.

One expects that many more constraints and interesting properties of superconformal field theories can be derived from the knowledge of this complete classification of unitary superconformal multiplets. Here, we present a theorem which directly follows from the classification.
\begin{customthm}{}\label{thm}
In the discrete space of superconformal field theories in six dimensions, there does not exist any sequence of theories that converges to a free theory.
\end{customthm}
The theorem only requires a very weak notion of convergence. The convergence of a sequence of theories to a free theory means that given any positive number $\widehat\Delta$, for each operator in the free theory with scaling dimension $\Delta < \widehat\Delta$, there exists an operator in each theory in the sequence, such that their scaling dimensions converge to $\Delta$.\footnote{In the usual discussion of weak coupling limits, the notion of convergence (of a sequence of theories) is much stronger. For instance, it may require the convergence of the entire spectrum of local operators and all the operator product expansion (OPE) coefficients.}

\subsubsection*{Review of superconformal representation theory}

To prove this theorem, let us first briefly review the procedure of constructing superconformal multiplets and some important facts about their classification \cite{Buican:2016hpb,Cordova:2016emh}. Superconformal primaries are operators that are annihilated by all superconformal $S$-supercharges. They form representations of the Lorentz and R-symmetry algebras. A highest-weight state is a superconformal primary that is also annihilated by the raising generators of the Lorentz and R-symmetry algebras. Given a highest weight state, a superconformal multiplet can be constructed by successive actions of the Poincar\'{e} $Q$-supercharges and the lowering generators of the Lorentz and R-symmetry algebras. 

Assuming unit norm for the highest-weight state, the norm of the descendant states are fixed by the scaling dimension, spins, and R-charges of the highest-weight state. In unitary theories, the norm of a state must be positive. Given the spins and R-charges of a generic highest-weight state, the absence of negative and zero norm states gives a lower bound on the scaling dimension $\Delta$,
\ie\label{eqn:unitaritybound}
\Delta> \Delta_{\cal A}.
\fe
Highest-weight states with scaling dimensions below the unitarity bound \eqref{eqn:unitaritybound} generically have negative norm descendants. However, when the scaling dimension $\Delta$ is equal to $\Delta_{\cal A}$ or some special values below $\Delta_{\cal A}$, the descendant states in the superconformal multiplet have only positive or zero norms. The zero norm states form a smaller representation of the superconformal algebra, and can be consistently decoupled, which makes the original multiplet shorter. The shortened multiplets are referred as {\it short multiplets}.

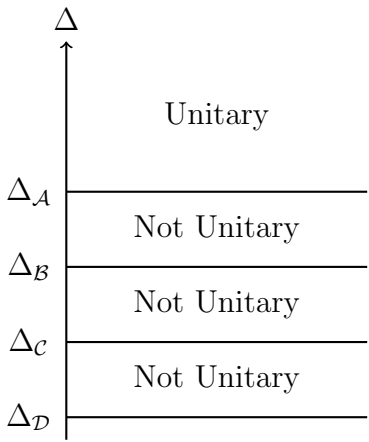
\begin{figure}[h]
\centering
\begin{tikzpicture}
\draw [->,thick](0,-.3) -- (0,5);

\draw [thick](0,3)--(4,3);
\draw [thick](0,2)--(4,2);
\draw [thick](0,1)--(4,1);
\draw [thick](0,0)--(4,0);

\draw (0,5.3) node{$\Delta$};

\draw (-.5,3) node{$\Delta_{{\cal A}}$};
\draw (-.5,2) node{$\Delta_{{\cal B}}$};
\draw (-.5,1) node{$\Delta_{{\cal C}}$};
\draw (-.5,0) node{$\Delta_{{\cal D}}$};

\draw (2,4) node{Unitary};
\draw (2,2.5) node{Not Unitary};
\draw (2,1.5) node{Not Unitary};
\draw (2,.5) node{Not Unitary};

\end{tikzpicture}
\caption{The scaling dimensions of long and short multiplets.}\label{fig:multiplets}
\end{figure}

The short multiplets of the six dimensional superconformal algebras are classified into ${\cal A}$, ${\cal B}$, ${\cal C}$, ${\cal D}$-types. The ${\cal A}$-type short multiplets sit at the lower bound of the continuum of long multiplets (the multiplets without shortening). The ${\cal B}$, ${\cal C}$, ${\cal D}$-type multiplets are isolated from the long multiplets by finite gaps in the scaling dimensions, and are referred as {\it isolated short multiplets}. The scaling dimensions of long and short multiplets are depicted in Figure~\ref{fig:multiplets}.

Consider a long multiplet with scaling dimension $\Delta=\Delta_{\cal A}+\epsilon$. In the $\epsilon\to0$ limit, the long multiplet decomposes into an ${\cal A}$-type short multiplet together with an ${\cal A}$, ${\cal B}$, ${\cal C}$, or ${\cal D}$-type short multiplet.
The decomposition rules (often called recombination rules) are studied in detail in \cite{Buican:2016hpb,Cordova:2016emh}. The short multiplets that do not participate in the decomposition are called {\it absolutely protected short multiplets}, in the sense that they are protected under any deformation that preserves the superconformal symmetry. Of course, all the absolutely protected short multiplets are isolated short multiplets. One key fact that we will use in proving our theorem is the following \cite{Buican:2016hpb,Cordova:2016emh}.
\begin{customlem}{}\label{lem}
In six dimensions, all conserved currents (including higher spin conserved currents) are in absolutely protected short multiplets.
\end{customlem}

\subsubsection*{Proof of main theorem}

Now, we are ready to prove our main theorem. Assume the contrary, that there is a sequence of superconformal field theories that converges to a free theory. If we label the theories in the sequence by positive integers $k=1,2,3,\cdots$, then the $k$-th theory must contain operators ${\cal O}_{s}^{(k)}$ (with spin labeled by $s$), such that the scaling dimensions $\Delta_{s}^{(k)}$ converge in the $k \to \infty$ limit to the scaling dimensions of the higher spin conserved currents in the free theory, {\it i.e.}
\ie
\lim_{k\to\infty}\Delta_{s}^{(k)} = 4+s.
\fe
This contradicts the above lemma. Hence, the theorem follows. 
 \hfill$\square$

\bigskip

Before discussing the implication of our theorem on the existence of actions, we discuss generalizations of our theorem to lower dimensions and applications to the bulk duals.

\subsubsection*{Generalization to lower dimensions}
In five dimensions, all conserved currents are in absolutely protected short multiplets \cite{Buican:2016hpb,Cordova:2016emh}. Hence, our theorem directly extends to five dimensions. There is no analogous statement in dimensions lower than five, since there are higher spin conserved currents in ${\cal A}$-type short multiplets, which are not absolutely protected (nor isolated). Indeed, the weak coupling limits in the space of four dimensional ${\cal N}=1,2,4$ theories and three dimensional ${\cal N}=1,\cdots, 8$ theories are realized by super-Yang-Mills theories and Chern-Simons matter theories, respectively.

There is a refined statement for the case of ${\cal N}=3$ supersymmetry in four dimensions: the genuine ${\cal N}=3$ theories -- {\it i.e.} ${\cal N}=3$ theories without extra SUSY-current multiplets -- do not admit weak coupling limits.\footnote{An ${\cal N}=4$ theory viewed as an ${\cal N}=3$ theory always contains extra SUSY-current multiplets, since an ${\cal N}=4$ stress tensor multiplet decomposes in to an ${\cal N}=3$ stress tensor multiplet and a complex conjugate pair of ${\cal N}=3$ extra SUSY-current multiplets.}  If such a weak coupling limit exists, it would be a sequence of genuine ${\cal N}=3$ theories converging to either a genuine ${\cal N}=3$ free theory or an ${\cal N}=4$ free theory. The former can be ruled out by the CPT theorem, since CPT invariant ${\cal N} = 3$ free theories must have enhanced ${\cal N} = 4$ supersymmetry, and hence fail to be genuine.\footnote{We thank Himanshu Raj for pointing this out.} The later can be ruled out by the fact that extra SUSY-current multiplets are absolutely protected \cite{Cordova:2016emh}.


%

\subsubsection*{Implication for string theory}

Under the AdS/CFT correspondence, the higher spin conserved currents are dual to higher spin gauge fields, and the higher spin non-conserved operators correspond to massive higher spin fields in the bulk theory. Our theorem thus implies that there does not exist a massless limit for the massive higher spin fields. If the bulk theory is a string theory, then there does not exist a tensionless limit, since the higher spin string modes become massless in the tensionless limit of the strings.\footnote{We thank Babak Haghighat for a discussion on this point.} Hence, we arrive at the following corollary.
\begin{customcoro}{}
Superstring theories on ${\rm AdS}_6$ and ${\rm AdS}_7$ do not have tensionless limits.
\end{customcoro}
An open problem is to understand this corollary purely from the bulk string theory perspective. On all ${\rm AdS}_6$ vacua of massive type IIA supergravity \cite{Brandhuber:1999np,Bergman:2012kr,Passias:2012vp} and the recently found ${\rm AdS}_6$ vacua of IIB supergravity \cite{Lozano:2012au,DHoker:2016ujz,DHoker:2016ysh,DHoker:2017mds,DHoker:2017zwj}, the string theories are not weakly coupled, as the dilaton always diverges at certain points on the internal manifold/orbifold.\footnote{We thank Martin Fluder for a discussion on this point.} The absence of a  tensionless limit may be due to the potentially nontrivial renormalization of the string tension. However, there are ${\rm AdS}_7$ vacua of massive type IIA supergravity \cite{Apruzzi:2013yva}, where the dilaton is everywhere finite. It would be interesting to understand the obstruction to taking the tensionless limit in these theories.

\subsubsection*{Towards the non-existence of actions}

Suppose that there exists an action for a superconformal field theory in five or six dimensions. Then physical observables can be computed by the path integral
\ie\label{eqn:PI1}
\int [D\Phi]\, e^{i S[\Phi]}\,(\cdots),
\fe
where $\Phi$ collectively denotes the fundamental fields used in the construction of the action functional $S[\Phi]$.\footnote{We allow the possibility that the action $S[\Phi]$ is non-local.} There can be gauge transformations acting on the fundamental fields, under which the action functional is invariant up to possible $2\pi$ shifts. By introducing a positive integer $k$ into the path integral
\ie\label{eqn:PI2}
\int [D\Phi]\, e^{i k S[\Phi]}\,(\cdots),
\fe
one generates a sequence of theories.\footnote{Without loss of generality, we assume the parameter $k$ cannot be absorbed by a field redefinition. If otherwise, the action functional would be a homogeneous function and can be written as
\ie
S[\Phi]=r\widetilde S[\widetilde \Phi],
\fe
where $r$ is a constant mode and $\widetilde \Phi$ denotes the fundamental fields that are independent of $r$. One obtains a new action by integrating out the constant mode $r$.}
In general, the parameter $k$ flows under renormalization.\footnote{For example, consider SU($N_c$) Yang-Mills theory in four dimensions coupled to Dirac fermions in the fundamental representation of the gauge group. When the flavor number $N_f$ is in the conformal window, the theory flows to the Caswell-Banks-Zaks fixed point with Yang-Mills coupling $g_{\text{\tiny YM}}=g_*$ determined by the $N_c$ and $N_f$ \cite{Caswell:1974gg,Banks:1981nn}. Along the renormalization group flow, the path integral of this theory can be written in the form \eqref{eqn:PI2} with the overall coupling $k=g^2_* / g_{\text{\tiny YM}}^2$.} Nevertheless, to proceed, we assume that $k$ does not flow, so that the path integral \eqref{eqn:PI2} defines a sequence of superconformal field theories.

In the large $k$ limit, the path integral can be evaluated by a saddle point approximation. More precisely, by expanding around the saddle point of the path integral, the action takes the form as a quadratic kinetic term plus interaction terms that are suppressed by $1/k$.\footnote{We assume that the dominate saddle point $\Phi_*$ of the action $S[\Phi]$ is non-degenerate, {\it i.e.} $S''[\Phi_*] \neq 0$.}  In general, the quadratic kinetic term may have higher derivatives. For example, the most general quadratic kinetic term for a scalar field $\sigma$ is
\ie\label{eqn:scalarAction}
\int d^d x\, \partial_\mu\sigma \Box^\delta \partial^\mu\sigma,
\fe
where $\Box=\partial_\mu\partial^\mu$. When $\delta = 0$, the theory is a unitary free theory. When $\delta$ is a positive integer, the theory is a non-unitary free theory, since the scaling dimension of the scalar $\sigma$ violates the unitarity bound. The stress tensor of this theory can be constructed from a bilinear of the scalar field $\sigma$, which takes the schematic form\footnote{The explicit formulae for $\delta = 1$ and 2 can be found in \cite{Osborn:2016bev}.}
\ie\label{eqn:T}
T_{\mu\nu} \sim \sigma   \overleftrightarrow\partial_\mu\overleftrightarrow\partial_\nu \Box^\delta\sigma.
\fe
For other values of $\delta$, the theory does not have a stress tensor, since the construction \eqref{eqn:T} gives a non-local operator.
In other words, the scalar field $\sigma$ is a generalized free field. Our theorem forbids the sequence \eqref{eqn:PI2} to converge to the theory of the action \eqref{eqn:scalarAction} with $\delta=0$. One can run similar arguments for the quadratic kinetic terms that consist of fermions, vector fields, or tensor fields. Our theorem implies that the sequence \eqref{eqn:PI2} can only converge in the $k\to\infty$ limit to a non-unitary free theory or a theory of generalized free fields.

If the sequence converges to a non-unitary free theory, then there must exist a positive integer $k_0$ such that the theories defined by the path integral \eqref{eqn:PI2} with $k> k_0$ are non-unitary.\footnote{A non-unitary free theory contains non-unitary operators of scaling dimensions that violate the unitary bound, which is denoted schematically as $\Delta\ge \Delta_*$. Given a non-unitary operator ${\cal O}$ of scaling dimension $\Delta<\Delta_*$, the convergence of a sequence of operators ${\cal O}_k$ to $\cal O$ means that there exists a positive integer $k_0$ such that for all $k>k_0$, $|\Delta_k - \Delta| < |\Delta_* - \Delta|$, where $\Delta_k$ denotes the scaling dimension of ${\cal O}_k$. This in turn implies that the theories in the sequence with $k>k_0$ are all non-unitary.
} In particular, this implies that the classical theory is non-unitary, but the theory with $k=1$ (which can be viewed as the classical theory under quantum corrections) is unitary. We find this rather unlikely.

If the sequence converges to a theory of generalized free fields, then in the $k\to\infty$ limit, the stress tensor must decouple, or equivalently the conformal central charge $C_T$ must diverge. This typically happens in the large $N$ limit of some matrix-like or vector-like models, where the number of fundamental fields diverges in the $N\to\infty$ limit.\footnote{For example, consider the critical O($N$) vector model in three dimensions with the action
\ie
\int d^3 x\left[{1\over 2}(\partial_\mu \phi^i)^2+{1\over 2}\sigma \phi^i \phi^i-{1\over 4g}\sigma^2\right].
\fe 
The infrared fixed point is achieved by sending $g\to \infty$. 
In the $N\to\infty$ limit, after integrating out $\phi^i$, the auxiliary field $\sigma$ becomes a generalized free field with $C_T=3N/2$ and the kinetic term
\ie
{N\over 16}\int d^3x\,\sigma {1\over \sqrt{\Box}}\sigma.
\fe}
However, the number of fundamental fields in the path integral \eqref{eqn:PI2} does not increase with $k$.
Hence, it is unlikely that the sequence of the path integrals \eqref{eqn:PI2} really converges to a theory of generalized free fields.

\section*{Acknowledgments}

I am grateful to Clay Cordova, Martin Fluder, Michael Geracie, Babak Haghighat, Ying-Hsuan Lin, Markus Luty, David M. Ramirez, Mukund Rangamani, Shu-Heng Shao, Yifan Wang, and Xi Yin for helpful discussions, and to Martin Fluder, Ying-Hsuan Lin and Xi Yin for comments on the first draft. I thank the Bootstrap 2018 at California Institute of Technology and Yau Mathematical Sciences Center at Tsinghua University for hospitality during the course of this work. I am supported by U.S. Department of Energy grant DE-SC0009999 and funds from the University of California.

\bibliography{refs} 
\bibliographystyle{JHEP}

\end{document}